\documentclass[aip,apl]{revtex4-1}

\usepackage[utf8]{inputenc}
\usepackage{graphicx}
\usepackage{amsmath}
\usepackage{amsfonts}
\usepackage{amssymb}
\usepackage{dcolumn}
\usepackage{bm}

\usepackage{textgreek} 
\newcommand{\um}{{\textmu}m }
\newcommand{\us}{{\textmu}s }
\newcommand{\ulmin}{{\textmu}l/min }

\begin{document}

	\title{Real-time sensing of flowing nanoparticles with electro-opto-mechanics}
	
	\author{Jeewon Suh$^1$, Kewen Han$^1$, Christopher W Peterson$^2$, and Gaurav Bahl$^{1\ast}$\\
		\footnotesize{$^1$ Department of Mechanical Science and Engineering} \\
		\footnotesize{$^2$ Department of Electrical and Computer Engineering} \\
		\footnotesize{University of Illinois at Urbana-Champaign, Urbana, Illinois 61801, USA}\\
		\footnotesize{$^\ast$ To whom correspondence should be addressed; bahl@illinois.edu} \\
	}
	
	\date{\today}
	
	\begin{abstract}
		High-Q optical resonators allow label-free detection of individual nanoparticles through perturbation of optical signatures but have practical limitations due to reliance on random diffusion to deliver particles to the sensing region. 
		We have recently developed microfluidic optomechanical resonators that allow detection of free-flowing particles in fluid media with near perfect detection efficiency, without requiring labeling, binding, or direct access to the optical mode. 
		Rapid detection of single particles is achieved through a long-range optomechanical interaction that influences the scattered light spectra from the resonator, which can be quantified with post-processing.
		Here, we present a hybrid electromechanical-optomechanical technique for substantially increasing the bandwidth of these optomechanofluidic sensors, enabling real-time operation. 
		The presented system demonstrates temporal resolution of better than 20~\us (50,000 events/second) with particle sensing resolution down to 490 nm, operating in the air without any stabilization or environmental control. 
		Our technique significantly enhances the sensing capabilities of high-Q optical resonators into the mechanics domain, and allows extremely high-throughput analysis of large nanoparticle populations.
	\end{abstract}
	
	\maketitle

	
	\vspace{12pt}

	
	\begin{figure}[b]
		\centering
		\includegraphics[width=\textwidth]{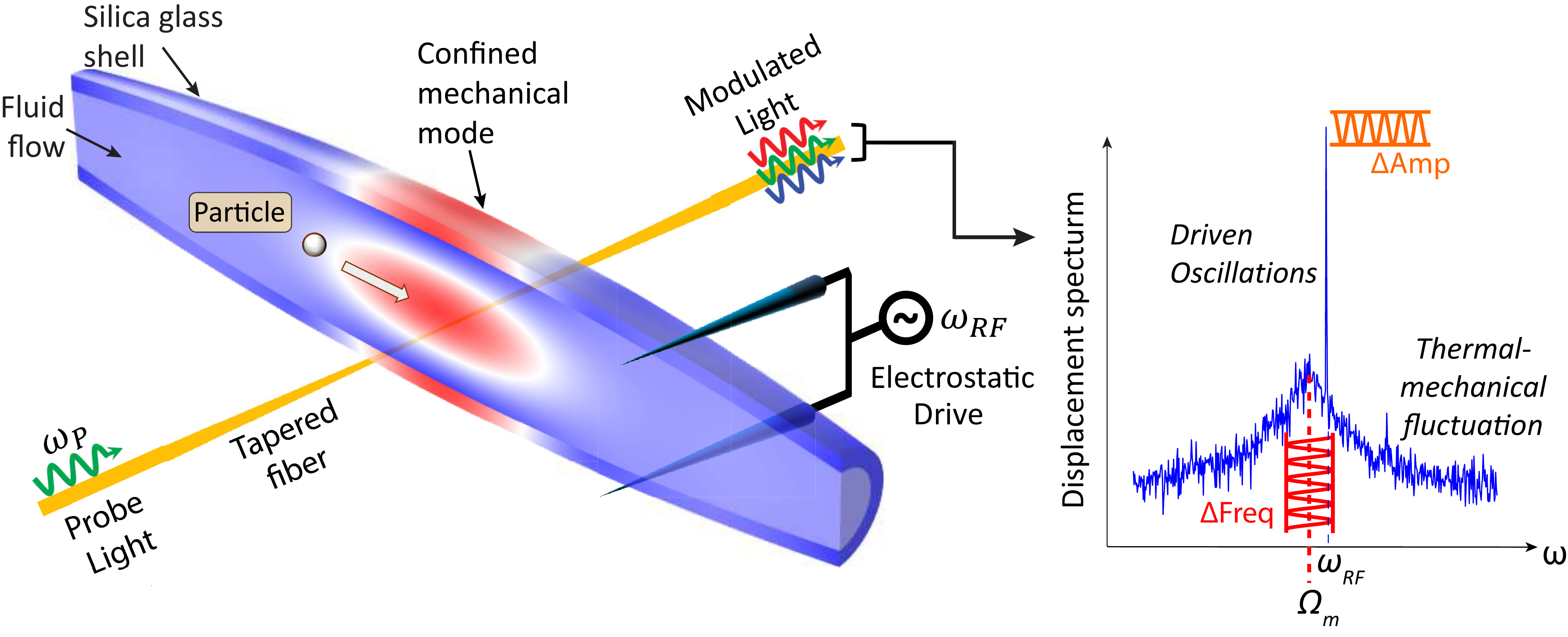}
		\caption{
			\textbf{Principle of opto-mechano-fluidic detection of flowing particles in a micro-capillary resonator.}
			The resonator simultaneously confines optical whispering gallery modes and breathing vibrational modes in the same physical region, resulting in strong optomechanical coupling. A tapered optical fiber allows probing of the resonator optical modes. Forward scattered light exhibits sidebands at the mechanical frequency, allowing the mechanical spectrum to be measured through self-heterodyne detection at a forward photodetector.
			The mechanical mode is driven by an applied electrostatic drive force at $\omega_{RF}$ by means of two 100 \um wire electrodes. Frequency perturbations of the natural vibrational frequency $\Omega_m$ caused by flowing particles are translated to amplitude modulation of the optical probe signal at $\omega_{RF}$.
		}
		\label{fig:principle}
	\end{figure}

	Resonant optical biosensors have enabled extremely sensitive label-free detection of single micro- and nano-particles in fluid samples \cite{Zhu2007, Fan2008, Vollmer2008, Lu2011, Kim2011, Dantham2012, Dantham2013, Baaske2014, Arnold2015, Yu2016,Zhu2010}.
	However, these sensors generally rely on the random diffusion and adsorption of particles within the sensing volume, which only enables quantification of a small fraction of the analyte. Furthermore, mechanical properties of these particles such as density, viscoelastic dissipation, and elastic modulus, which do not couple directly to optical fields, cannot be quantified by these techniques. 
	In this context, the utilization of optomechanical coupling \cite{Rokhsari2005,Carmon2005,Kippenberg2005,Liu2013} may allow both long-range sensing of particles using phonons without adsorption near the optical mode, and the extraction of mechanical parameters, using only optical measurements \cite{Han2014_1,Zhu2014}.
	This optomechanical technique has been recently implemented using opto-mechano-fluidic resonators (OMFRs) \cite{Bahl2013,HyunKim2013, Kim2014} which are composed of a hollow fused silica microcapillary and support co-located optical and mechanical modes. 
	The large optomechanical coupling between the modes of OMFRs has been previously used to develop sensors for the density \cite{HyunKim2013} and viscosity \cite{Han2014_1} of bulk fluid samples. 
	Recently, we have demonstrated that the vibrational modes existing in OMFRs cast a nearly perfect 'phonon net' that may also be used to quantify the mechanical properties of individual particles that rapidly flow through the OMFR microchannel, without requiring adsorption \cite{Han2016}. In this work, we introduce a new experimental method for significantly improving the detection speed and signal-to-noise ratio of these particles measurements through the use of electro-opto-mechanical transduction. 
	
	Our preliminary work \cite{Han2016} demonstrated that the center frequency and linewidth of the OMFR vibrational modes are sensitive to the mechanical parameters of flowing particles. These measurements were performed by optically extracting the spectrum of the thermal-mechanical fluctuations of the vibrational mode. However, since this vibrational noise signal is close to the noise floor of the measurement apparatus, the measurements require spectral analysis, averaging, and curve fitting of the optical and electronic signals. The instrumentation limit of these measurements is ultimately set by the capabilities of the hardware with which spectra can be evaluated. In the present work, we demonstrate that processing-intensive extraction of the entire vibrational spectrum is not necessary and that it is instead sufficient to quantify the mechanical transfer function at a single frequency. Optical detection of this mechanical response to an applied force is shown to enable real-time measurements of the vibrational mode frequency fluctuations during particle transit events. Our experiments showcase a system operating in the air without any environmental controls, that has a measurement noise floor of 490 nm (particle diameter) and can sample transits as fast as 20 \us.
	Our experiments are a major step towards practical real-time flow cytometry analysis of the mechanical properties of individual particles.

	
	\vspace{12pt}

	\begin{figure}[bh!]
		\centering
		\includegraphics[width=0.35\textwidth]{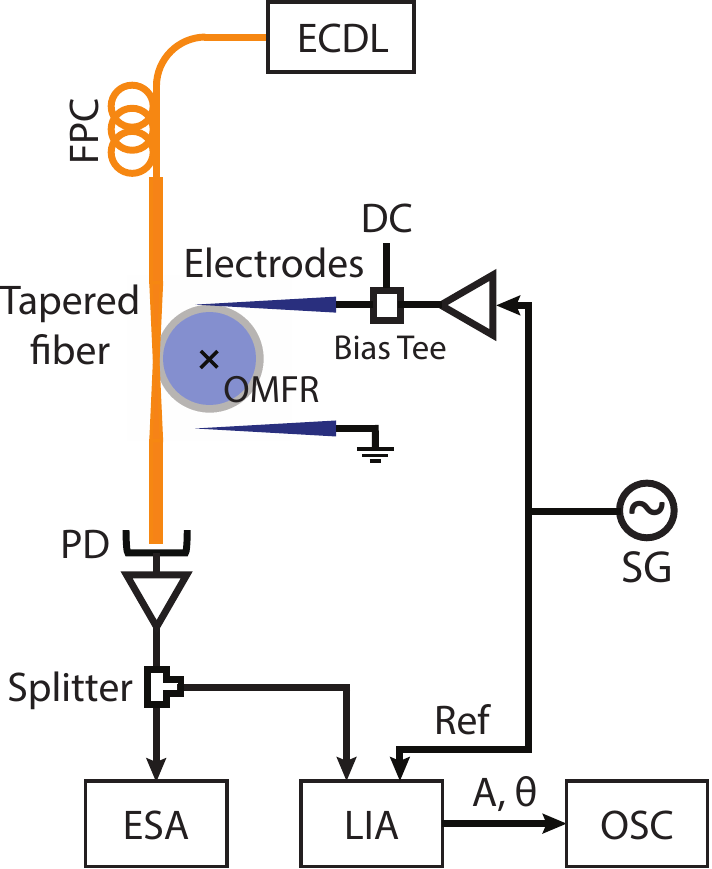}
		\caption{
			\textbf{Real-time measurement system:} 
			The OMFR's high-Q optical whispering-gallery modes are probed by a fiber-coupled external cavity diode laser (ECDL) through a tapered optical fiber. The fiber is put in contact with the OMFR to stabilize the coupling against vibration and long-term coupler drift. A signal generator (SG) is used to provide the RF stimulus at $\omega_{RF}$ to the two electrodes that are also placed in contact with the OMFR to ensure direct mechanical actuation. The bias tee and DC voltage are used to enhance the electrostatic drive force. The two amplifiers mitigate the influence of the RF electronic feedthrough signal.
			A forward photodetector (PD) converts the optomechanically modulated light to an electronic signal, carrying information on the vibrational spectrum. This signal is monitored by an electrical spectrum analyzer (ESA) and a lock-in amplifier (LIA). The lock-in outputs (amplitude A and phase $\theta$) are monitored in real-time using an oscilloscope (OSC).
		}
		\label{fig:schematic}
	\end{figure}

	The fabrication of OMFRs has previously been reported in \cite{Bahl2013,Han2014_J}. Briefly, these microcapillary resonators are fabricated from fused silica capillaries (Polymicro Technologies TSP-700850) of 850 \um outer diameter that are adiabatically drawn under laser heating used for softening the material. The diameter of an OMFR can be locally varied by modulating the laser power during the drawing process. Microcapillaries having microbottle geometry with 60-70 \um outer diameter can be easily produced by this method, and support simultaneous confinement of optical and mechanical modes in the regions of the highest diameter. One end of the device is connected to a syringe pump, using which analyte solutions can be flowed through at the desired rate. The OMFRs are vertically oriented during experiments to prevent particles from settling due to gravity. 
	
	Fig.~\ref{fig:schematic} shows a schematic of our experimental setup. We employ a telecom laser probe (1520-1570 nm) to observe the OMFR's optical whispering-gallery modes (WGMs) through coupling with a tapered optical fiber. 
	The OMFR is placed in contact with the tapered optical fiber to minimize fluctuations of the taper-resonator coupling due to ambient vibration or by vibration induced due to fluid flow. 
	Both thermal-mechanical fluctuations (and driven oscillations) of the co-located mechanical modes temporally modulate the optical WGMs, resulting in modulation of the probe laser signal propagating forwards in the fiber. The displacement spectrum of these mechanical modes can thus be quantified using the beat note formed between the probe laser and its modulation sidebands on a photodetector, with the help of an electronic spectrum analyzer (ESA). Perturbations of the mechanical mode caused by flowing particles can be detected rapidly by this method.
	Typical eigenfrequencies of these mechanical vibrations span 20-40 MHz. 
	While this detection system fundamentally has a very high bandwidth, the measurement rate is limited by the capability and complexity of the ESA. %
	This measurement challenge is compounded by the fact that the thermal-mechanical fluctuation signal is generally very close to the ESA measurement noise floor, thus rendering real-time analysis difficult.
	Thus, quantification of the amplitude and frequency parameters of the mechanical vibration usually requires post-processing of the measured spectra.

	In this work, we implement an electro-opto-mechanical technique \cite{Lee2010,Taylor2012} that considerably improves both the measurement rate and signal-to-noise ratio by directly actuating the mechanical mode using an electrostatic drive force. The resulting optomechanical modulation signal can be easily measured 50 - 60 dB above the electronic noise floor, as seen in the spectrum example presented in Fig.~\ref{fig:principle}. 
	The electrostatic actuator is implemented using two 100 \um diameter electrodes placed on either side of the OMFR, with a large signal applied at $\omega_{RF}$ near the mechanical resonance. One of the electrodes is placed in contact with the OMFR to ensure direct mechanical actuation. 
	A lock-in amplifier (Stanford Research Systems model SR844) is used to monitor in real-time the electrical-to-optical transfer function of the system at the frequency $\omega_{RF}$. Frequency shifts of the vibrational mode result in both amplitude and phase perturbations of this transfer function, with bandwidth only subject to the lock-in time constant. Prior to and after each experiment, we calibrate the amplitude response of the full system as a function of frequency, which is then used for frequency-tracking the phonon mode.

	
	\vspace{12pt} 	
	

	\begin{figure}[ph!t]
		\centering
		\includegraphics[width=0.82\textwidth]{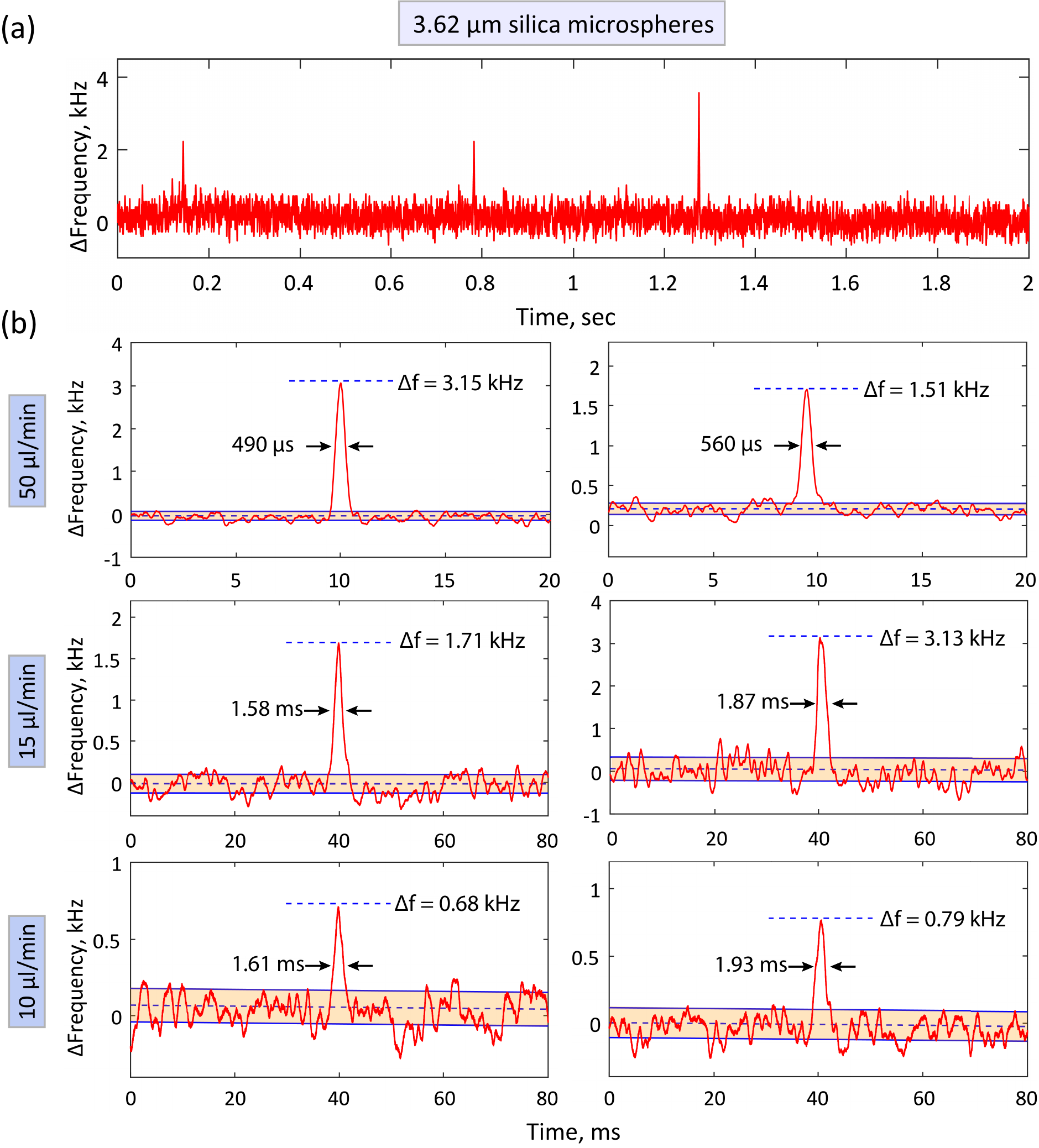}
		\caption{
			\textbf{Real-time measurements of particle transits:} 
			(a) Successive transits of 3.62 $\pm$ 0.09 \um silica particles stand out above the background frequency noise. 
			(b) Zoomed views of various transits measured at 10, 15, and 50 \ulmin flow rate. Each measurement quantifies the magnitude of frequency shift and the transit speed of the particle. The transit time is calculated as the full-width at the average frequency shift between the local peak and the local background average. The shaded box in each data set represents the mean and $\pm$ one standard deviation.
		}
		\label{fig:particle}
	\end{figure}

	In this experiment, we use an OMFR with 70 \um outer diameter, 50 \um inner diameter, and perform measurements using a 24.26 MHz mechanical mode. 
	The analyte solution is composed of monodisperse 3.62 $\pm$ 0.09 \um silica particles (Cospheric SiOMS-1.8 3.62 \um) mixed in water. The particles are significantly diluted to ensure transits that are clearly separated in time.

\begin{figure}[b]
		\centering
		\includegraphics[width=0.85\textwidth]{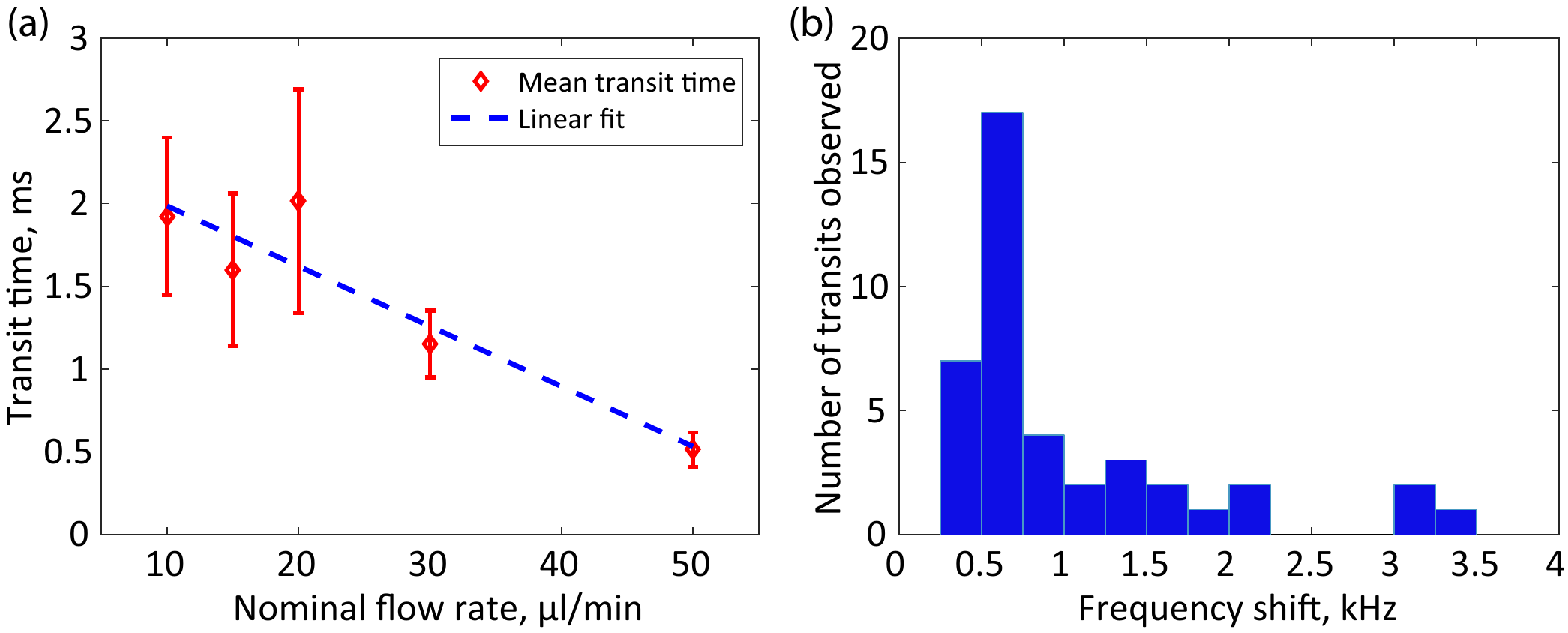}
		\caption{
			\textbf{Analysis of transit time and frequency shift statistics:} 
			(a) The average transit time is presented from multiple observed transits at various flow rates. 
			The error bars represent one standard deviation. The transit time is linearly proportional to the nominal flow rate, as expected. 
			(b) We generate a histogram of all measured transits into 250 Hz bins for the 10, 15, and 50 \ulmin flow rates. While most transits cluster in the 0.5 - 0.75 kHz range, there is deviation since individual particles follow different streamlines within the flow.}
		\label{fig:collection}
	\end{figure}

	In Fig.~\ref{fig:particle}(a), we show that frequency shifts of the vibrational mode associated with individual particle transits can be clearly observed above the background noise fluctuation in real-time, without needing any post processing. %
	Fig.~\ref{fig:particle}(b) shows a sampling of individual particle transits at flow rates of 10 \ulmin, 15 \ulmin, and 50 \ulmin; zoomed in to show detail and the extremely high rate of measurement. 
	These observations are made using the same OMFR and the same vibrational mode, allowing an estimation of transit speed and the frequency perturbation caused by each individual particle. 
	The frequency shift associated with each transit is different even though the particles are nominally monodisperse. This variation is well known from our previous work \cite{Han2016} and occurs as a function of the local properties of the phonon mode where the transit occurs. Additionally, background noise levels between all the presented transits in Fig.~\ref{fig:particle}(b) are similar, and only appear different due to the different vertical scaling of each figure.
	Using this technique, we are able to detect particle transits with timescales as short as 490 \us, which is presently only limited by the achieved syringe pump flow rate. This timescale corresponds to a measurement throughput exceeding 1000 particles/second, which is 40x faster than our previous result \cite{Han2016}.
	As anticipated, the particle transit times decrease (Fig.~\ref{fig:collection}(a)) as the nominal flow rate at the syringe pump is increased.

	
	We gathered observations of 41 transits of the analyte particles at flow rates of 10, 15, and 50 \ulmin as shown in Fig.~\ref{fig:collection}(b). 
	The vibrational mode frequency shift associated with these transits shows clustering around 0.5-0.75 kHz with significant variation beyond this range. This spread is not explained by the 5\% diameter variation in particles that the manufacturer has characterized.
	The large variations in sensitivity are thus most likely caused because of inadequate control over particle trajectory through the OMFR, i.e. particles follow different streamlines in the liquid flow. The sensitivity of the system with respect to particle location has been previously discussed in \cite{Han2016}. It is thus imperative for future experiments to implement hydrodynamic focusing (i.e. sheath and core flow) which is a standard technique used in flow cytometry. This will also help make the possible consistent measurement of particle populations and comparisons between individual particle groups in mixed analytes.
	
	\vspace{12pt}
	
	The limit of measurement achieved by this system is set by both random and systematic frequency fluctuations in the mechanics. Generally, acquiring measurements by averaging over a longer time interval does reduce the effect of rapid fluctuations and improves the noise floor of any measurement system, provided that long-term drift is not significant. On the other hand, longer sample acquisition time can reduce the measurement bandwidth, and therefore an optimization of acquisition interval is necessary. 
	The optimized averaging time and corresponding sensing limit can be determined by means of Allan deviation analysis \cite{Allan1966, Allan1997}, which is a two-sample deviation measurement often used to identify noise sources in an oscillator. 
	Quantification of the Allan deviation of the oscillator frequency measurement as a function of the acquisition period (i.e. averaging time), can serve as a guide to selecting the best averaging time to obtain the best noise floor in our measurements.

		
		\begin{figure}[t]
			\centering
			\includegraphics[width=0.75\textwidth]{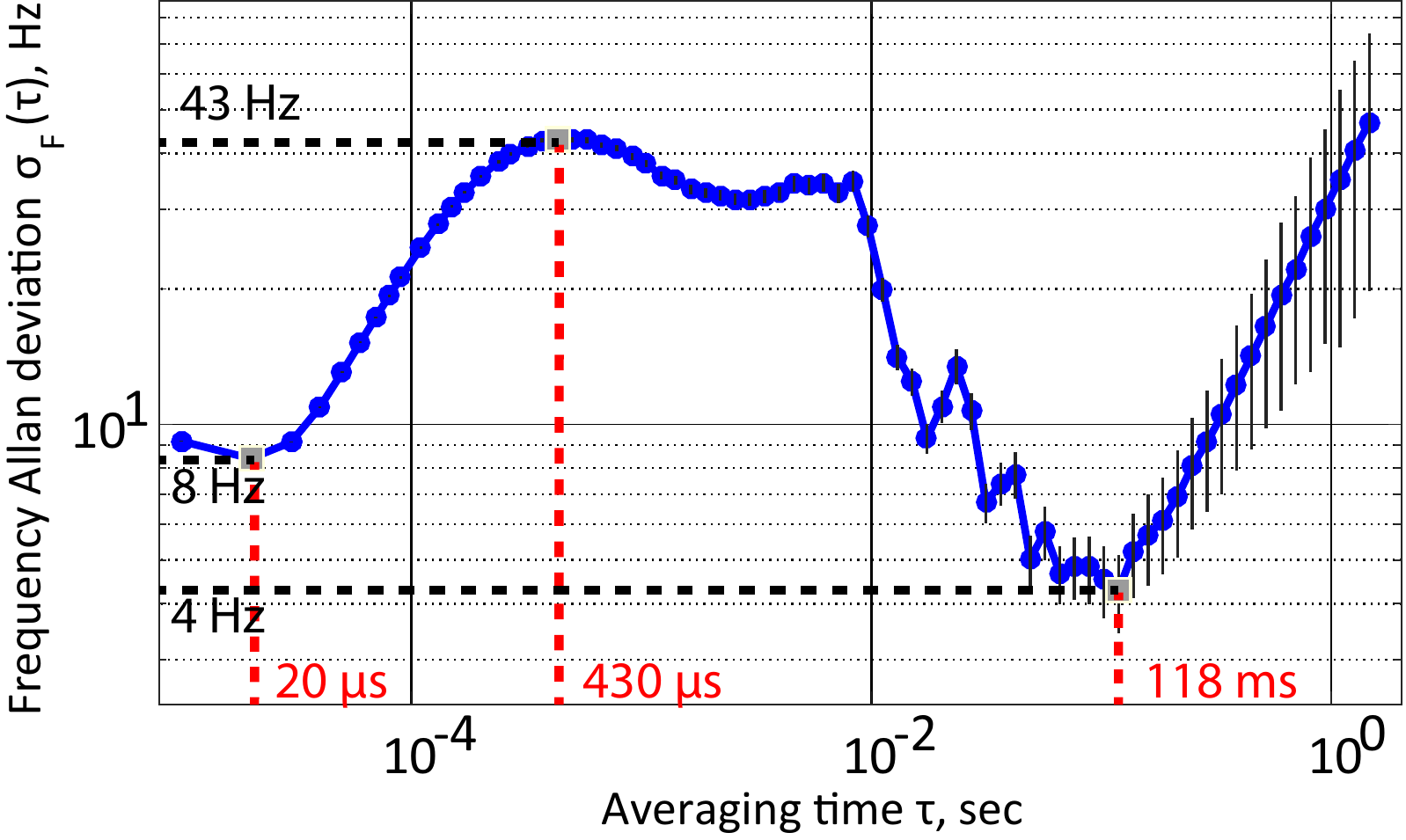}
			\caption{
				\textbf{Quantification of the measurement noise floor:} We perform Allan deviation analysis for frequency fluctuations that are measured with only liquid flowing through the OMFR. We indicate the frequency Allan deviation for a few key averaging times that represent important timescales in this system (see text).}
			\label{fig:allan}
		\end{figure}

	In Fig.~\ref{fig:allan} we present Allan deviation measurements of the mechanical frequency of the OMFR mode, without any particles present in the flowing carrier liquid (water). These measurements are taken without environmental protection or feedback stabilization, indicating the potential for further improvement.
	We see that the uncertainty is minimized for averaging time around 120 ms, with long-term drift affecting the measurement for longer averaging times. 
	However, since we aim to measure particle transits times at or below than 400 \us, the system cannot afford time-averaging of the frequency data to this extent. 
	At fast timescales, a minimum Allan deviation of 8~Hz is measured when using nearly the full bandwidth of this system ($\tau = 20$ \us, equaling 50,000 events/second).
	This Allan deviation quantification allows us to estimate the detection limit of the OMFR sensor, by comparison of the noise floor with calibration particle measurements. We first assume that hydrodynamic focusing will ultimately reliably allow us to access the highest sensitivity region of the vibrational mode.
	Presently, the greatest experimentally observed vibrational frequency shift for the 3.6 \um particle population is $\Delta f_{max} = 3.13$ kHz, which is likely obtained by particles transiting a high-sensitivity region of the vibrational mode. 
	However, since higher sensitivity particle trajectories may exist, we may only use this 3.13 kHz detection as a lower-bound estimate of the highest achievable sensitivity.
	Using the measured 8 Hz Allan deviation for averaging time of 20 \us, a volume ratio analysis obtains the particle size noise floor as 490 nm.
	The current OMFR system is thus very close to the regime where single viruses (typically 100 nm) may be detectable.

	
	\vspace{12pt}

	Presently, our real-time system for detecting free-flowing nanoparticles in solution operates at a single frequency and cannot characterize the complete mechanical spectrum. Even so, the particle density and compressibility properties can be derived from these simple measurements \cite{Han2016}. 
	Ultimately, it is trivial to extend this system for performing multi-frequency measurements of the optomechanical response. 
	In the near future, the uncertainties arising from vibration, temperature, pressure \cite{Han2014_2,Yang2016} may also be mitigated through such multi-frequency measurements performed on several mechanical modes simultaneously.
	Finally, positive feedback could be used to dynamically boost the sensitivity of the system \cite{Taylor2012} but with a sacrifice of measurement bandwidth.

	OMFR devices are composed of transparent silica glass, which allows for simultaneous optical measurements on the individual flowing particles that are typically tagged with fluorescent dyes, in the same manner as traditional flow cytometry.
	However, a transformative new capability can be imparted to flow cytometers through the addition of real-time measurement of the mechanics of individual nanoparticles without binding or tagging.
	The simultaneous extraction of optical and mechanical responses for single bioparticles could provide new degrees of information that have not previously been accessible.

	\begin{acknowledgments}	
	
	Funding for this research was provided through the National Science Foundation (NSF) grants ECCS-1408539 and ECCS-1509391, and the Air Force Office for Scientific Research (AFOSR) Young Investigator grant FA9550-15-1-0234. The authors would also like to thank Alan Luo for continuing experimental assistance.
	
	\end{acknowledgments}

	\bibliography{fastsensing}

\end{document}